\documentclass[preprint,5p]{elsarticle}

\pdfoutput=1
\usepackage{mathrsfs}
\usepackage{amsmath}
\usepackage{setspace}

\usepackage[pdftex]{hyperref}
\hypersetup{
      colorlinks=true,
      linkcolor=blue,
      citecolor=blue
}

\begin{document}

\begin{frontmatter}
	
\title{Fast neutron background characterization with the Radiological Multi-sensor Analysis Platform (RadMAP)}

\author[LBL,USMA]{John R. Davis\corref{cor1}}
\ead{john.davis@usma.edu}

\author[SNL]{Erik Brubaker}

\author[LBL]{Kai Vetter}

\cortext[cor1]{Principal corresponding author}
\address[LBL]{Lawrence Berkeley National Laboratory, Berkeley, CA, USA}
\address[USMA]{The United States Military Academy, West Point, NY, USA}
\address[SNL]{Sandia National Laboratories, Livermore, CA, USA}

\begin{abstract}

In an effort to characterize the fast neutron radiation background, 16 EJ-309 liquid scintillator cells were installed in the Radiological Multi-sensor Analysis Platform (RadMAP) to collect data in the San Francisco Bay Area. 
Each fast neutron event was associated with specific weather metrics (pressure, temperature, absolute humidity) and GPS coordinates. The expected exponential dependence of the fast neutron count rate on atmospheric pressure was demonstrated and event rates were subsequently adjusted given the measured pressure at the time of detection.   
Pressure adjusted data was also used to investigate the influence of other environmental conditions on the neutron background rate. 
Using National Oceanic and Atmospheric Administration (NOAA) coastal area lidar data, an algorithm was implemented to approximate sky-view factors (the total fraction of visible sky) for points along RadMAP’s route.  Three areas analyzed in San Francisco, Downtown Oakland, and Berkeley all demonstrated a suppression in the background rate of over 50\% for the range of sky-view factors measured. This effect, which is due to the shielding of cosmic-ray produced neutrons by surrounding buildings, was comparable to the pressure influence which yielded a 32\% suppression in the count rate over the range of pressures measured. 

\end{abstract}


\end{frontmatter}

\section{Introduction}\label{sec:Intro}

Neutron detection is a key component of mobile wide-area search for nuclear materials or devices. Neutrons provide a sensitive and specific signature of special nuclear material (SNM). 
All SNM sources emit neutrons as a result of spontaneous fission events, with Pu materials emitting on the order of 10$^{5}$ n/(s${\cdot}$kg). In contrast, highly enriched uranium (HEU) with 90\% $^{235}$U and 10\% $^{238}$U content emits less than 4 n/(s${\cdot}$kg). Therefore, quantities of Pu are typically the focus of passive fast neutron detection systems searching for SNM~\cite{Runkle_SecSNM}. 

As in any detection application, background radiation limits the ability to detect hidden sources with confidence. An appealing feature of fast neutron detectors for SNM searches is that the background is mostly constant and relatively low, especially when compared to the much more abundant and variable gamma-ray background. Nevertheless, depending on the scenario (time of exposure, source to detector distance, shielding and surrounding materials), detection of SNM neutrons above background levels with high confidence can be limited by systematic variability in the background rates. Understanding the factors that influence the naturally occurring radiation field is therefore crucial for confident detection.

Characterization of background radiation on a mobile platform has many advantages. An extensive variety of weather, altitude, and other environmental conditions are attainable on a daily basis. The Radiological Multi-sensor Analysis Platform (RadMAP) and its 16 organic liquid scintillator cells were utilized by Lawrence Berkeley National Laboratory (LBNL) to collect fast neutron (500 keV to 8 MeV) background data throughout the San Fransisco Bay Area beginning in May 2012. The area is known for its many micro-climates, all of which are readily accessible to RadMAP's location at LBNL. A variety of structural conditions are also present. Long bridges of differing construction, tunnels, dense urban environments, and sparse rural areas are all located within a 30 mile range of Berkeley. The terrain also offers the ability to take measurements from sea level to over 3800 feet (Mount Diablo in Contra Costa County). This paper presents the results of a comprehensive study of the environmental influences on the fast neutron background. Research demonstrated the significance of the altitude and atmospheric pressure influence on the count rate, as documented in literature. It was determined that applying a pressure adjustment to measured event rates improves background predictability by reducing systematic error contributions. After applying a pressure adjustment to the data, the effects of other weather metrics, solar weather, and surrounding structures were investigated. A study of the influence of the shielding provided by surrounding structures was conducted by computing the fraction of unobstructed sky visible from the mobile platform and comparing it with the measured count rate.  This study expands on recent work by Iyengar et al.~\cite{Iyengar_Urban_Det}.

\section{Experimental system}\label{sec:Exp_sys}

\subsection{RadMAP and detector specifications}\label{sec:RadMAP_Spec}

\begin{figure}[h!]
	\begin{spacing}{1.0}
		\centering
		\includegraphics[scale=0.26]{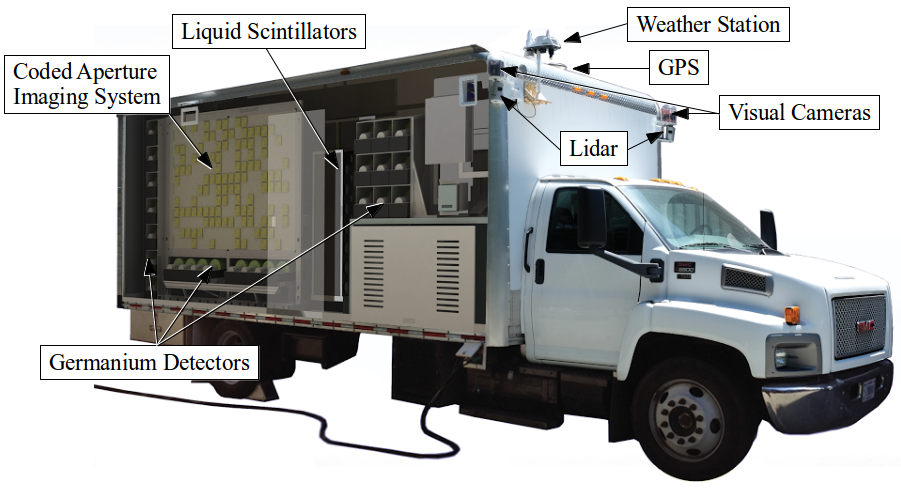}
		\caption{RadMAP schematic with a transparent rear side wall to illustrate positions of detection systems and sensors.}
		\label{fig:Sensors}
	\end{spacing}
\end{figure}

RadMAP, previously known as the Mobile Imaging and Spectroscopic Threat Identification (MISTI) system, was originally developed by the Naval Research Laboratory as a mobile gamma-ray source detection and localization platform. The platform is a General Motors 20 foot box truck with an on board generator to provide power to its detectors and sensors~\cite{Mitchell_MISTI_2}\cite{RadMAP_newpaper}. 
MISTI was transferred to LBNL and began acquiring data in the San Francisco Bay Area in November 2011. It was subsequently renamed RadMAP given its change of mission focus to background characterization and its additional suite of integrated sensors and detection capabilities. Following the transfer, the system was used primarily for gamma-ray background characterization and source detection studies. RadMAP began collecting fast neutron background data in May 2012 following the installation of the liquid scintillator cells.  Figure~\ref{fig:Sensors} shows a RadMAP schematic highlighting some of its detection systems and external sensors.

Between installation of the scintillators in May 2012 and December 2013, 37 mobile datasets are usable for neutron analysis.  Due to various maintenance issues, the truck was immobile during 2014. During this time, over 100 data sets of significant length (12-15 hours each) were collected from RadMAP's parking spot adjacent to Building 88 on LBNL. The large quantity of statistics collected during this time period enabled stationary measurements of various weather and geomagnetic conditions that influence the neutron background count rate.

\begin{figure}[h!]
	\begin{spacing}{1.0}
		\centering
		\includegraphics[scale=0.26]{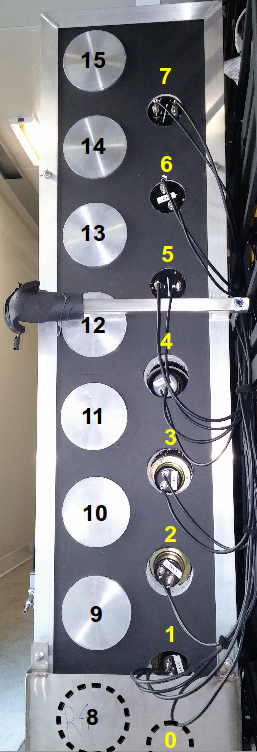}
		\caption{The 16 EJ-309 liquid scintillator array as viewed when facing the rear of the truck's cargo space. Detector channels 0 and 8 are not visible due to their location inside the leak containment safety structure.}
		\label{fig:Array}
	\end{spacing}
\end{figure}

The scintillator array provided by Sandia National Laboratories (SNL) in Livermore, California consists of 16 EJ-309 organic liquid scintillator cells for fast neutron detection. EJ-309 was designed for its pulse shape discrimination (PSD) characteristics or the ability to distinguish a neutron induced signal from a gamma-ray interaction.  The tail-to-total method was used for all PSD calculations in this experiment.  EJ-309 also is an outstanding candidate for field deployment due to its high flash point (291$^\circ$F), low vapor pressure and low chemical toxicity~\cite{EJman}, especially when compared to its predecessors such as the flammable solvent xylene. In RadMAP, each individual detector is oriented horizontally and stacked vertically in two columns of 8 detectors each as pictured and numbered in Fig.~\ref{fig:Array}.
Each cell is a 5 inch diameter by 5 inch long aluminum cylinder. The total active detection volume of the system is approximately 25 L. Seven detectors are coupled to a 5 inch Hamamatsu photomultiplier tube (PMT). The other nine are coupled to Photonis 5 inch PMTs. 
The PMTs are connected to two Struck SIS3320 250 MHz 12 bit digitizers, each with 8 channels. The digitizers are operated at 200 MHz, so a sample is recorded every 5 nanoseconds. The provided Struck data acquisition software is used to control the digitizers and collect raw data. Once the raw event signal pulses with associated timing information are recorded, the raw data is parsed into a usable format for data processing, PSD, and analysis.

Each identified neutron event was then associated with specific weather metrics (pressure, temperature, absolute humidity) and GPS coordinates for subsequent count rate analysis. A Davis Vantage Vue Wireless Weather Station~\cite{weatherstation} was used to collect all relevant weather data for the analysis.  The original GPS system installed on MISTI was a Magellan ADU5 which used four sensors, satellites, and ground-based stations to achieve down to 40 cm accuracy~\cite{oldgps_man}. However, the accuracy of the system was compromised when terrain interfered with the signal. A NovAtel Synchronous Position, Altitude and Navigation (SPAN) 
\linebreak GNSS/INS integrated GPS system~\cite{newgps_man} was installed in January 2012. The NovAtel system provides centimeter level accuracy and a data rate of 100 Hz. The incorporation of the inertial system provided more accurate positional information during periods of intermittent satellite reception.

\section{Altitude and pressure influence}\label{sec:Press}

In cosmic ray physics, the atmospheric depth (g/cm$^2$) is a measure of the path length traveled by a particle used to predict absorption. Atmospheric depth (g/cm$^2$) is the air density (g/cm$^3$) multiplied by the path length from the top of the atmosphere to a given location in cm. As measured by Pfotzer~\cite{Pfotzer}, at approximately 15 km altitude, the cascades of particles originating from a primary cosmic ray reach a maximum particle density. Below this point, also known as the Pfotzer point, an exponential decrease in the number of all particles in the cosmic ray induced shower is observed due to attenuation~\cite{Ziegler_CosmicRays}.   

The attenuation of cosmic rays below the Pfotzer point is commonly expressed in terms of an absorption length (also called the mean free path or attenuation factor) in units of g/cm$^{2}$. This unit may also be expressed in terms of a standard pressure unit such as mbar (1 mbar = 1.01972 g/cm$^{2}$). The absorption length differs for each type of particle depending on its mass, energy, and strength of interaction with the particles in the atmosphere~\cite{Ziegler_CosmicRays}. Equation~\ref{eq:N_I} is the exponential decay formula for computing an expected neutron intensity for any atmospheric depth~\cite{Desilets_scalefact}.
\begin{equation}\label{eq:N_I}
I_2 = I_1\text{exp}\left(\frac{x_1-x_2}{L}\right)
\end{equation}
$I_1$ is a measured neutron intensity recorded at depth $x_1$ and particle absorption length $L$, and $I_2$ is the expected intensity at depth $x_2$. The two depths and $L$ are in units of g/cm$^2$.

\begin{figure}[h!]
	\begin{spacing}{1.0}
		\centering
		\includegraphics[scale=0.46]{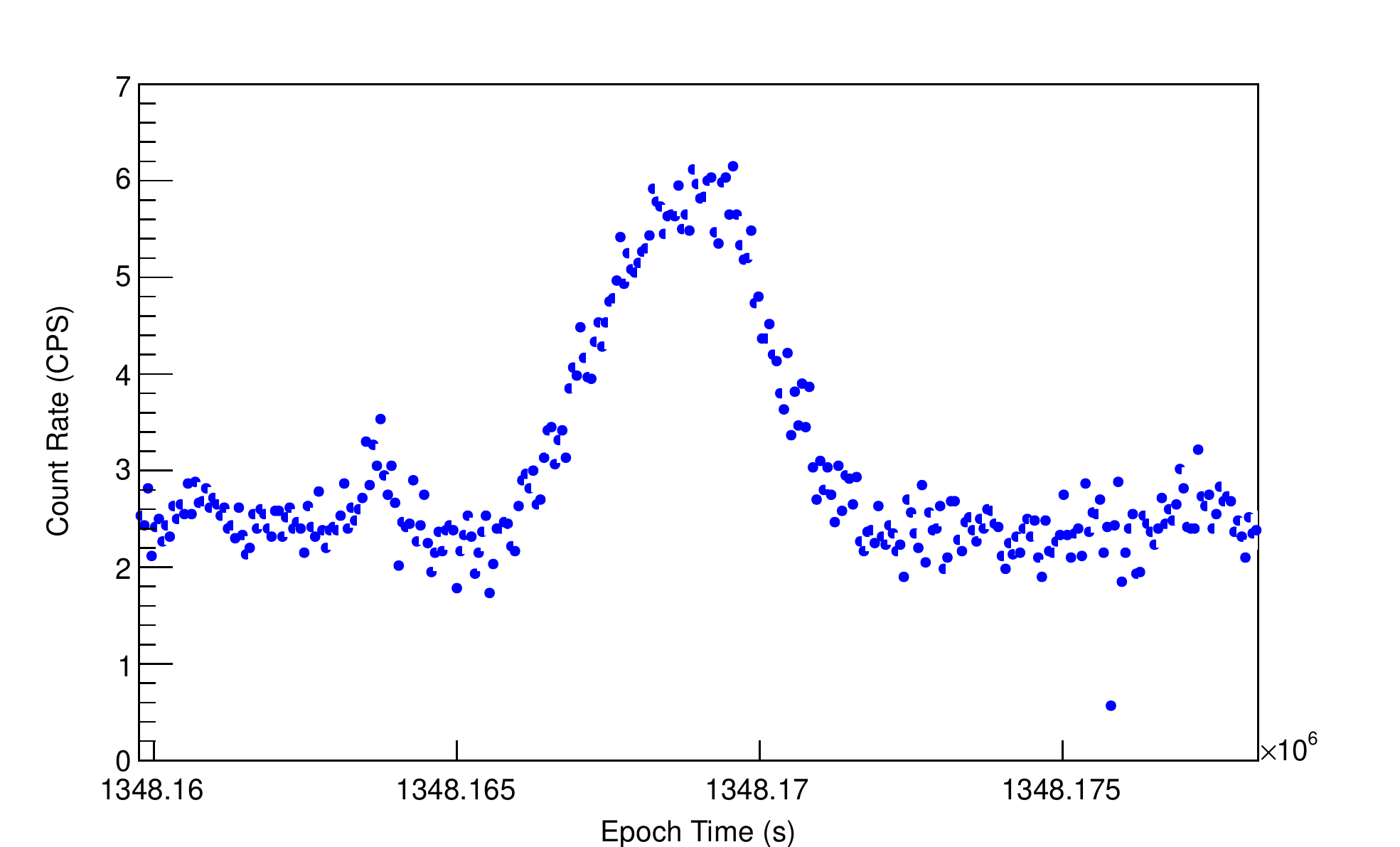}
		\caption{The one minute averaged count rates (CPS) for the duration of the Mount Diablo run.}
		\label{fig:DiabloRun}
	\end{spacing}
\end{figure}

Only one RadMAP run with the liquid scintillators data contains altitude data above 600 meters. On September 20, 2012 RadMAP made the round trip from LBNL to the peak of Mount Diablo (1,173 meters altitude). The changes in count rates when traveling along Grizzly Peak Boulevard, and during the ascent and descent of Mount Diablo is apparent in the run data represented in Fig.~\ref{fig:DiabloRun}.
The low count rate measured near the right end of the run corresponds to the time while traveling through the Solano Ave Tunnel.

\subsection{Altitude and pressure results}\label{sec:Press_result}

\begin{figure}[h!]
	\begin{spacing}{1.0}
		\centering
		\includegraphics[scale=0.32]{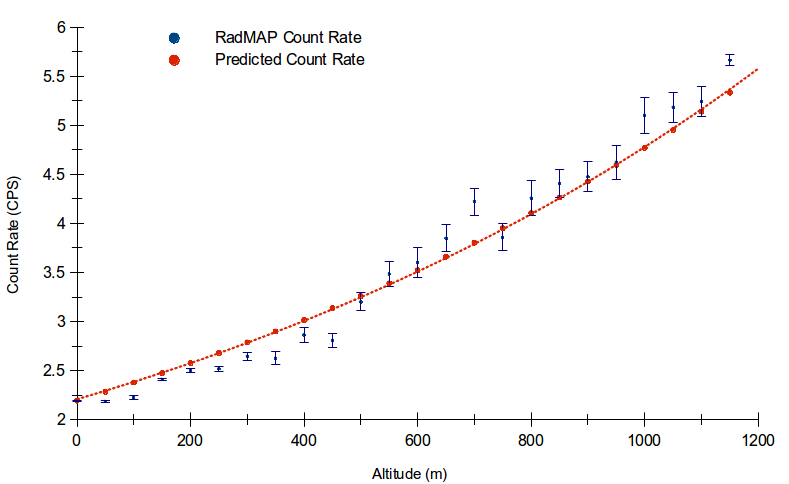}
		\caption{RadMAP mobile run observed and predicted neutron count rates using Ziegler's equations. The predicted rates are relative to the sea level measured count rate.}
		\label{fig:Pfotzer_CR}
	\end{spacing}
\end{figure}

An altitude count rate histogram was created for every RadMAP run combined to compare to the observations of Pfotzer. Applying Eq.~\ref{eq:N_I} and an average neutron absorption length, $L$, used by Ziegler~\cite{Ziegler_CosmicRays} of 148 g/cm$^2$, a Pfotzer curve predicted count rate for each altitude bin may be plotted for comparison to the observed rates. For the Pfotzer predicted count rates in Fig.~\ref{fig:Pfotzer_CR}, the measured mean rate (not adjusted for pressure) of 2.2 counts per second (CPS) at sea level (0 meters) was used for $I_1$. This effectively pins the red Pfotzer curve in Fig.~\ref{fig:Pfotzer_CR} to the measured (blue data point) count rate at 0 meters. Ziegler's equation~\cite{Ziegler_CosmicRays} for converting altitude to atmospheric pressure was also used to convert altitude measurements to an equivalent pressure or atmospheric depth in g/cm$^2$ for use in Eq.~\ref{eq:N_I}.

The predicted count rate, $I_2$, is determined for each altitude bin given its atmospheric depth, $x_2$, and the constant (altitude-independent) values for $I_1$, $x_1$, and $L$. Fairly good agreement between the predicted and measured values was obtained in Fig.~\ref{fig:Pfotzer_CR}.

\begin{figure}[h!]
	\begin{spacing}{1.0}
		\centering
		\includegraphics[scale=0.48]{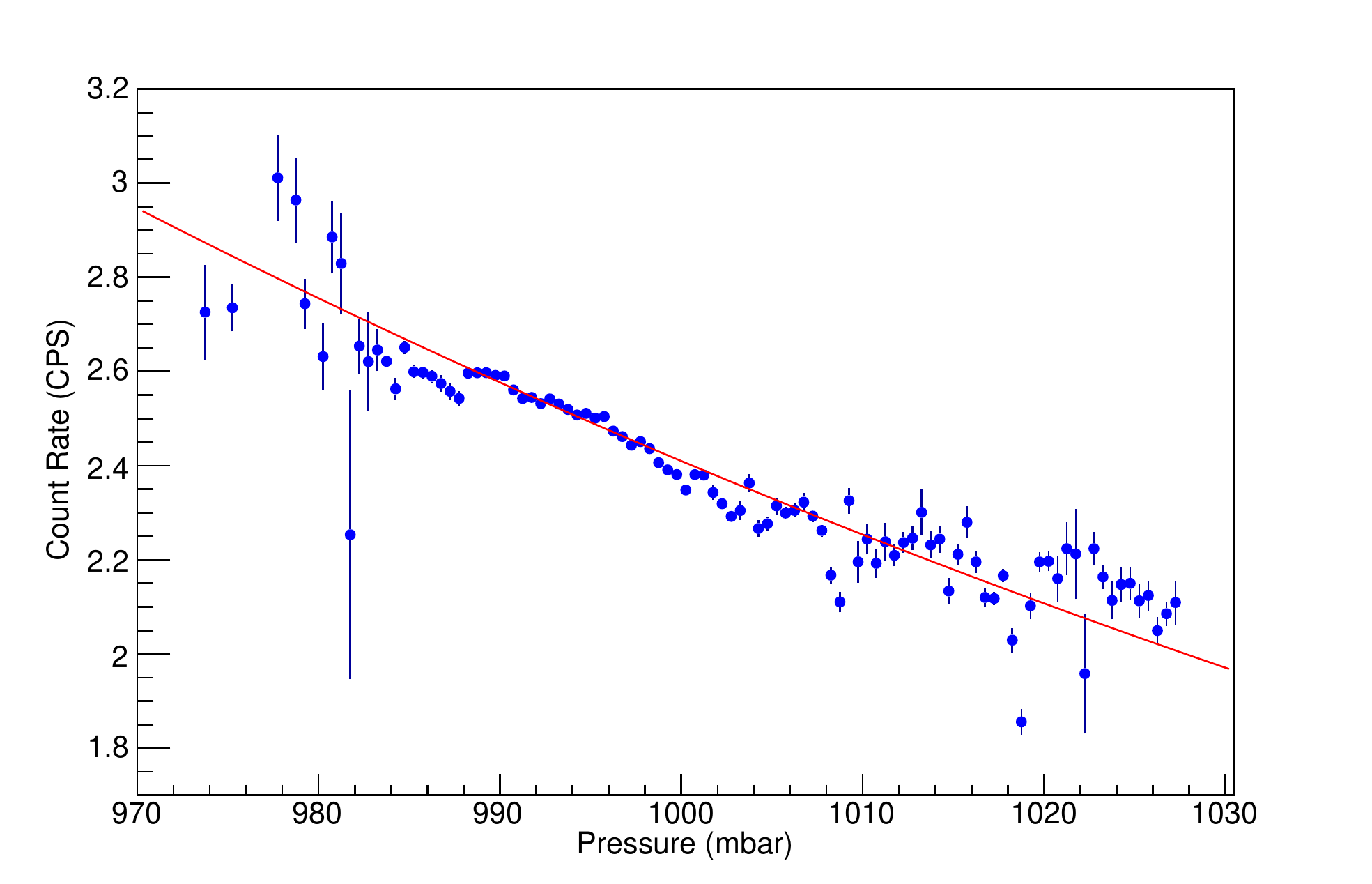}
		\caption{Count rates given atmospheric pressure measurements for all datasets fitted with the exponential function $y=\text{exp}(7.58-0.00670x)$. The area in the center with small error bars are due to the statistics collected during stationary data acquisition.}
		\label{fig:Press_CR_All}
	\end{spacing}
\end{figure}

As the count rate increased with increasing altitude above, the opposite relationship is expected for count rates at increasing pressures (as altitude increases, pressure decreases and there is less attenuation of cosmic ray neutrons). The result for the count rates at atmospheric pressure values averaged over all runs (including stationary measurements taken at Building 88) is plotted in Fig.~\ref{fig:Press_CR_All}.
A suppression of the count rate with increasing pressure occurred as predicted. The range in count rates measured from low to high pressure of 0.9 CPS (3 down to 2.1 CPS) corresponds to a significant suppression of 32\%.

\subsection{Pressure adjustment method}\label{sec:Press_adj}

At a constant altitude, a wide range of pressures may be observed within a few hours time. A significant variation of count rates is observed for the stationary data; therefore, for this analysis the count rates are adjusted based on atmospheric pressure, rather than on altitude. The fit parameters from the combined pressure data for all datasets (Fig.~\ref{fig:Press_CR_All}) were used to determine the appropriate count adjustment. The fitted count rate $C$ dependence on pressure $P$ is
\begin{equation}\label{eq:ROOT_exp}
C(P) = \text{exp}(7.58 - 0.00670 P\text{[mbar]})
\end{equation}
Confidence intervals for the fit parameters in Eq.~\ref{eq:ROOT_exp} are $(7.58\pm0.08)$ and $(0.00670\pm0.00008)$mbar$^{-1}$.

\begin{figure}[h!]
	\begin{spacing}{1.0}
		\centering
		\includegraphics[scale=0.48]{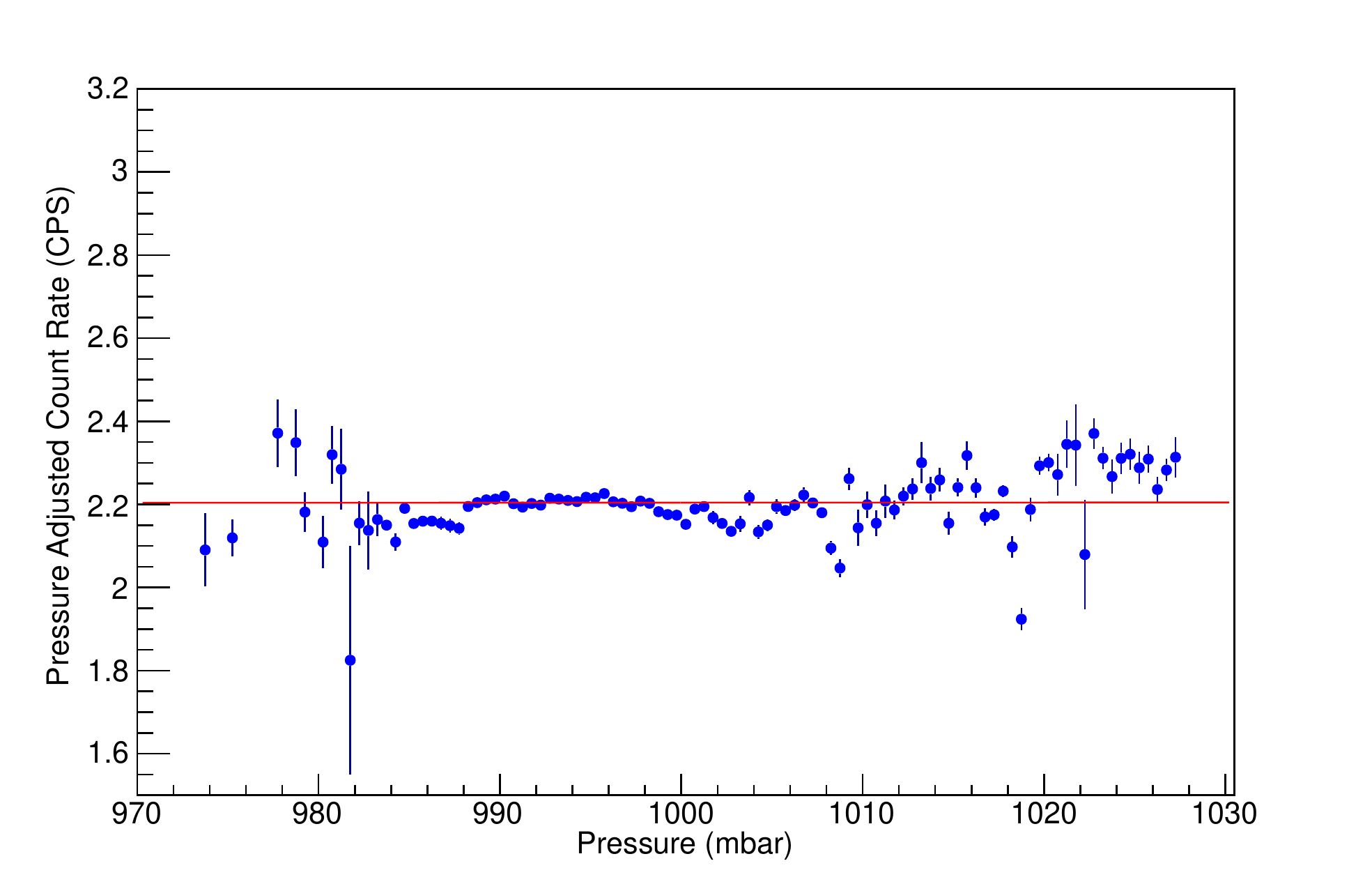}
		\caption{Pressure adjusted count rate histogram for all datasets with linear fit $y=(0.00001\pm0.00014) x+(2.2\pm0.1)$.}
		\label{fig:Adj_Press_CR}
	\end{spacing}
\end{figure}

\begin{figure}[h!]
	\begin{spacing}{1.0}
		\centering
		\includegraphics[scale=0.48]{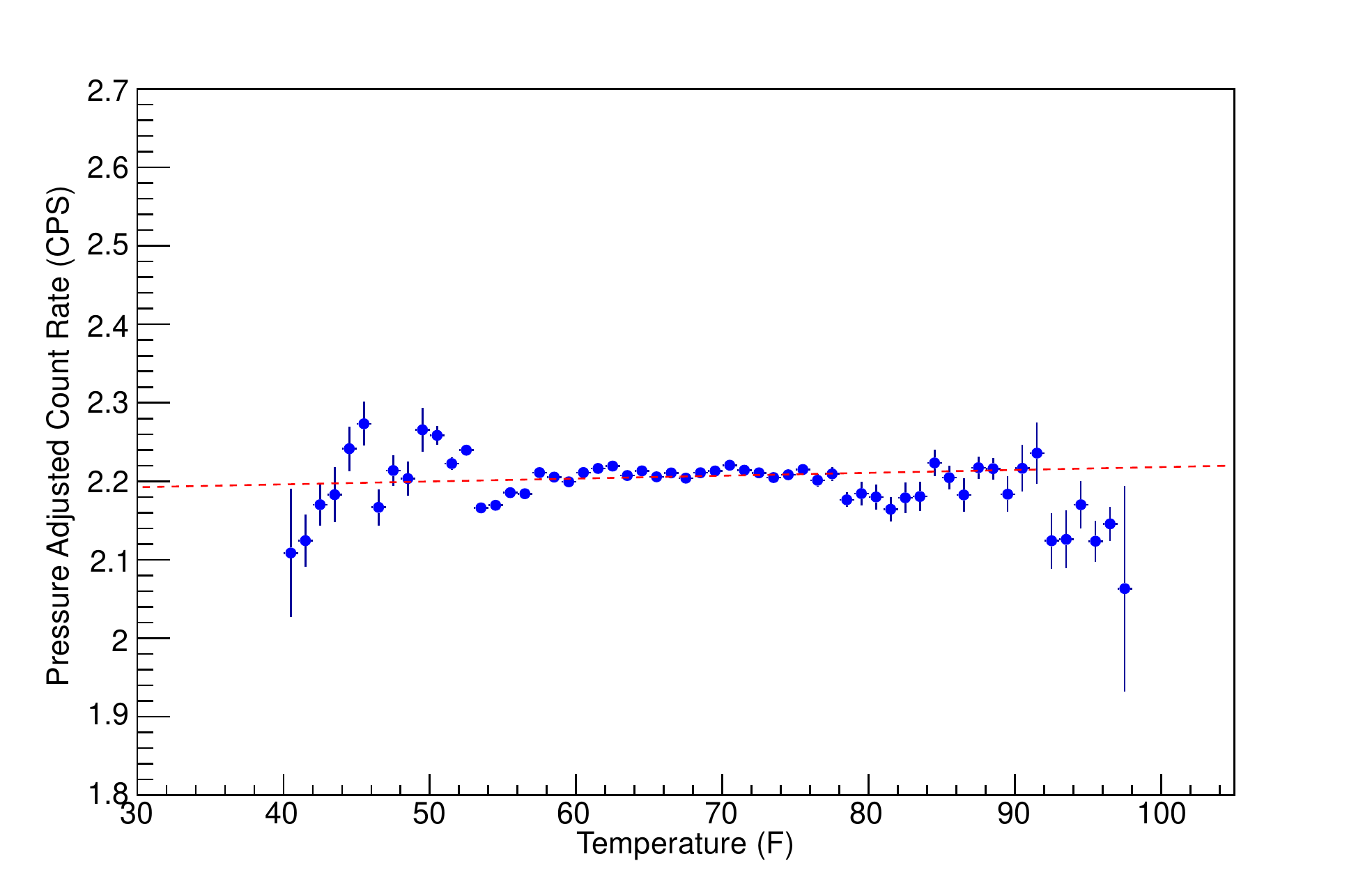}
		\caption{Pressure adjusted temperature count rate histogram. Equation for linear fit is $y=(0.0004\pm0.0001) x+(2.181\pm0.006)$.}
		\label{fig:Adj_Temp_CR}
	\end{spacing}
\end{figure}

\begin{figure}[h!]
	\begin{spacing}{1.0}
		\centering
		\includegraphics[scale=0.48]{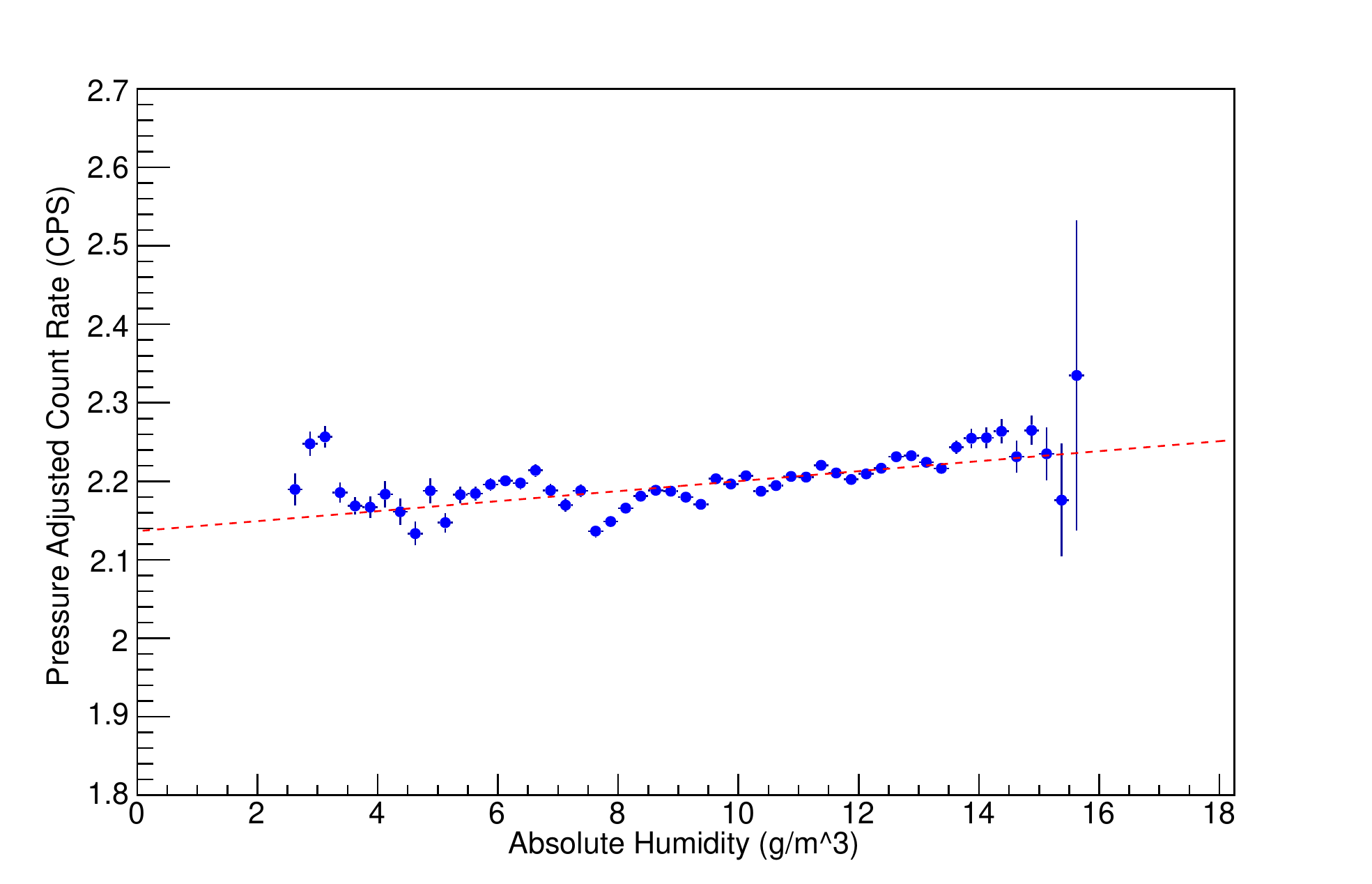}
		\caption{Pressure adjusted absolute humidity count rate histogram.  Equation for linear fit is $y=(0.0064\pm0.0003) x+(2.136\pm0.004)$.}
		\label{fig:Adj_Hum_CR}
	\end{spacing}
\end{figure}

In this paper, the pressure adjustment is applied by adjusting every neutron count to an equivalent number of counts (or fraction of a count) given the measured atmospheric pressure at the time of detection. The reference point used is one count at standard atmospheric pressure (1013.25 mbar). 
With the correction applied, if a neutron is detected at a measured pressure of 1013.25 mbar, its adjusted value will remain one count. However, if the pressure measurement is lower than standard pressure, the adjusted count value or weight is a fraction of one count. The adjusted weight of one count at measured pressure $x$ is set equal to the ratio of the expected count rates at standard pressure and at the measured pressure:
\begin{equation}\label{eq:ct_adj}
\text{Adjusted Count Weight} = \frac{C_\text{std}}{C_\text{meas}} = \frac{C(1013.25)}{C(x)}
\end{equation}
The resulting pressure count rate histogram with all counts adjusted in this manner should be a flat distribution with the best fit line located at the standard pressure or sea level mean count rate ($f(1013.25)=2.206$ CPS) as shown in Fig.~\ref{fig:Adj_Press_CR}.
Once the pressure dependency is effectively removed, various weather conditions and other environmental variables may be studied for residual correlations that may exist. For the energy range of neutrons measured in this study (500 keV to 8 MeV), no significant residual correlations was observed for temperature as depicted in Fig.~\ref{fig:Adj_Temp_CR}. There is a weak residual correlation observed with absolute humidity (Fig.~\ref{fig:Adj_Hum_CR}), but more data and analysis are needed to understand whether this relationship has predictive value or arises in this particular dataset from accidental correlations with other hidden variables. Higher absolute humidity results in more hydrogen atoms in the air and therefore more softening of the cosmogenic neutron spectrum. Any real slight increase in count rate with absolute humidity could be due to a greater number of high-energy neutrons downscattering into the detectable region than the number of neutrons originally within the detectable region that are then downscattered below the detection threshold.  

\subsection{Influence of pressure adjustment on count rate distributions}\label{sec:Press_CR_dist}

\begin{figure}[h!]
	\begin{spacing}{1.0}
		\centering
		\includegraphics[scale=0.48]{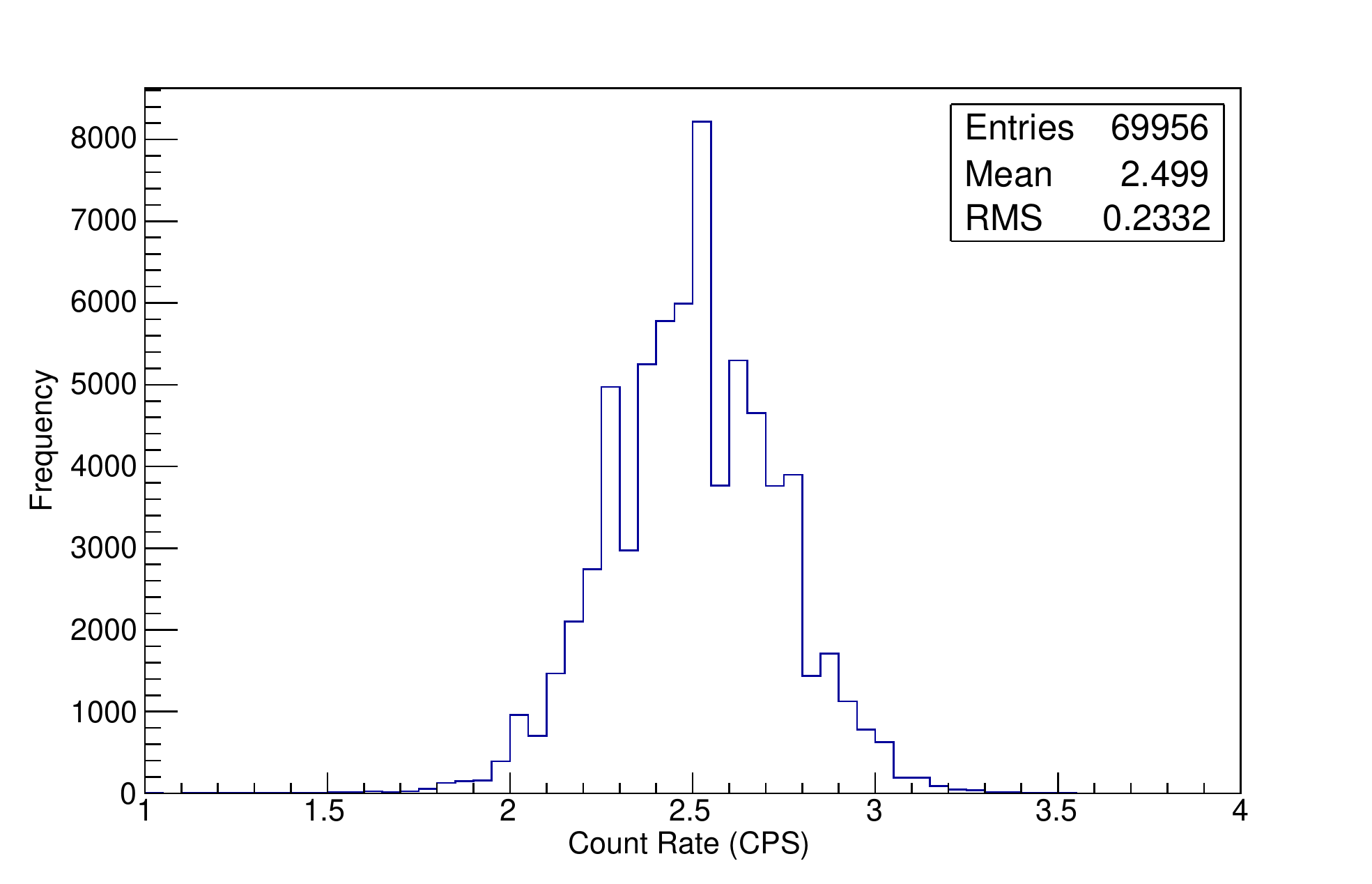}
		\caption{Unadjusted distribution of count rates for all RadMAP runs combined.}
		\label{fig:CR_Dist}
	\end{spacing}
\end{figure}

\begin{figure}[h!]
	\begin{spacing}{1.0}
		\centering
		\includegraphics[scale=0.48]{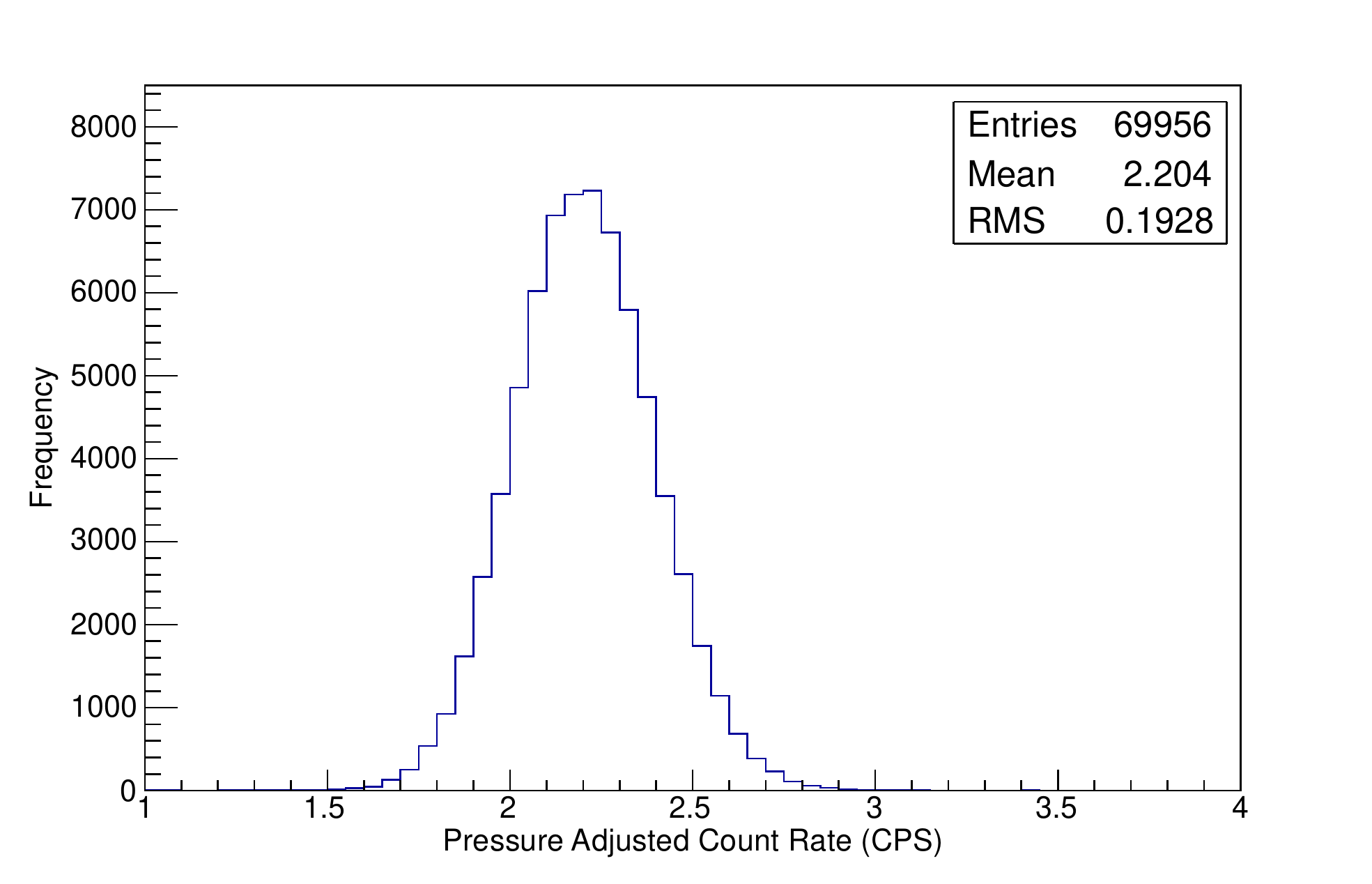}
		\caption{Pressure adjusted distribution of count rates for all RadMAP runs combined.}
		\label{fig:Adj_CR_Dist}
	\end{spacing}
\end{figure}

In this section, we estimate the value of the pressure correction in predicting a fast neutron background count rate. 
For every run, count rates are determined for every minute of data acquisition and the rate is filled into a count rate frequency histogram. In the absence of systematic variability, the result would be a Poisson distribution of observed rates centered about the mean or expected value. The distributions are described by their mean and root mean square (RMS) error as done for the unadjusted and pressured adjusted distributions in Fig.~\ref{fig:CR_Dist} and Fig.~\ref{fig:Adj_CR_Dist}, respectively.
Upon initial inspection, it is clear the mean shifts down to the sea level equivalent rate and the RMS is smaller for the pressure adjusted distribution. A certain quantity of the error is due to statistical uncertainty but there is also a contribution from systematic error from environmental variables. Given the count rate mean, $\mu_\text{rate}$, and RMS, $\sigma_\text{total}$, the statistical and systematic error contributions may be determined. 
The pressure adjusted distribution is scaled for calculation so the mean matches that of the unadjusted data. The RMS is also adjusted by the same factor to give the properly scaled error.  
Table~\ref{table:CRError} shows the results for each quantity in the error calculation.
\begin{table}[htbp]
	\centering
	\caption{Error calculations for count rate distributions for all runs with weather data combined.}
	\label{table:CRError}
	\includegraphics[scale=0.21]{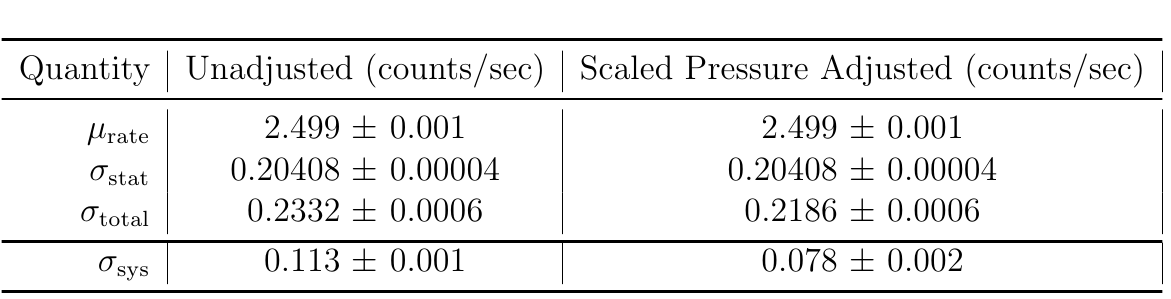}
\end{table}%

The total width of the unadjusted distribution is 0.2332 CPS which, after subtracting in quadrature the contribution from statistical uncertainty, gives a residual systematic uncertainty of 0.113 CPS. For the pressure adjusted distribution, the total and residual errors are 0.2186 CPS and 0.078 CPS, respectively. Thus by applying the pressure adjustment to this data, a reduction of 31\% in the systematic error is obtained. This narrowing of the distribution increases the understanding of the expected count rate and will result in greater sensitivity and specificity in detection of a source over background. Note that although the averaging time (here $60 \text{s}$) affects the statistical contribution to the width of the distribution, the systematic contribution is largely insensitive to that value as long as it is small compared to the timescale of the relevant systematic variability.

\section{Solar weather considerations and effects on neutron background count rate}\label{sec:solar_efx}

\subsection{Coronal Mass Ejections and Ground Level Enhancements}\label{sec:CME_GLE}

Two solar weather phenomena that may affect the measured cosmic-ray induced neutron count rate are Coronal Mass Ejections (CMEs) and Ground Level Enhancements (GLEs). CMEs are ejections from the sun which are often, but not always, accompanied by a solar flare. The component of CMEs that exist in the interplanetary field are called `ejecta' by Cane~\cite{Cane_CMEs} and are said to exhibit depressed plasma proton temperatures, bidirectional particle flows, and strong magnetic fields. The magnetic fields associated with ejecta and the interplanetary shock component formed by fast moving ejecta result in a net decrease in the cosmic ray count rate on earth due to deflections of primary galactic cosmic rays. 
Such decreases are also commonly referred to as Forbush decreases, crediting the first observations of such events made by Forbush in 1937~\cite{Cane_CMEs}. 

Ground Level Enhancements (GLEs) were not observed during the period of RadMAP's data acquisition. GLEs occur when a large quantity of particles ejected from the sun reach GeV energies that allow them to penetrate the earth's magnetic field. The result is an increase in neutron event rates in ground based detectors. Bieber, et al.~\cite{Bieber_GLE} and Plainaki, et al.~\cite{Plainaki_GLE} report results from a GLE event on January 20, 2005.

\subsection{The geomagnetic activity $K_p$ index}\label{sec:kp_idx}

The $K$ index is a metric that was introduced by Bartels in 1938 as a measure of geomagnetic field activity~\cite{Love_K_index}. It was designed to measure the local magnetic activity for a specific observatory given its well-understood quiet day activity levels, diurnal fluctuations, and other longer term variations. The result is an index that characterizes the strength of a geomagnetic storm. The $K$ index, which ranges from 0 to 9, is a quasi-logarithmic integer determined from the range of magnetic field intensities for a 3 hour time period. A planetary $K$ index may then be determined by taking the average $K$ index from 13 observatories (located between 44$^\circ$ and 60$^\circ$ N or S latitude)~\cite{NOAA_Kp} to further reduce local diurnal and seasonal influences. This average planetary $K$ index, known as the $K_p$ index, also ranges from 0 to 9 but in finer increments of 1/3. NOAA's National Geophysical Data Center regularly publishes the daily $K_p$ indices.

For analysis, the reported $K_p$ value is assigned to each detected neutron event to determine if there is a correlation between the geomagnetic field activity (as measured by the $K_p$ index) and the pressure-adjusted count rate. 
\begin{figure}[h!]
	\begin{spacing}{1.0}
		\centering
		\includegraphics[scale=0.48]{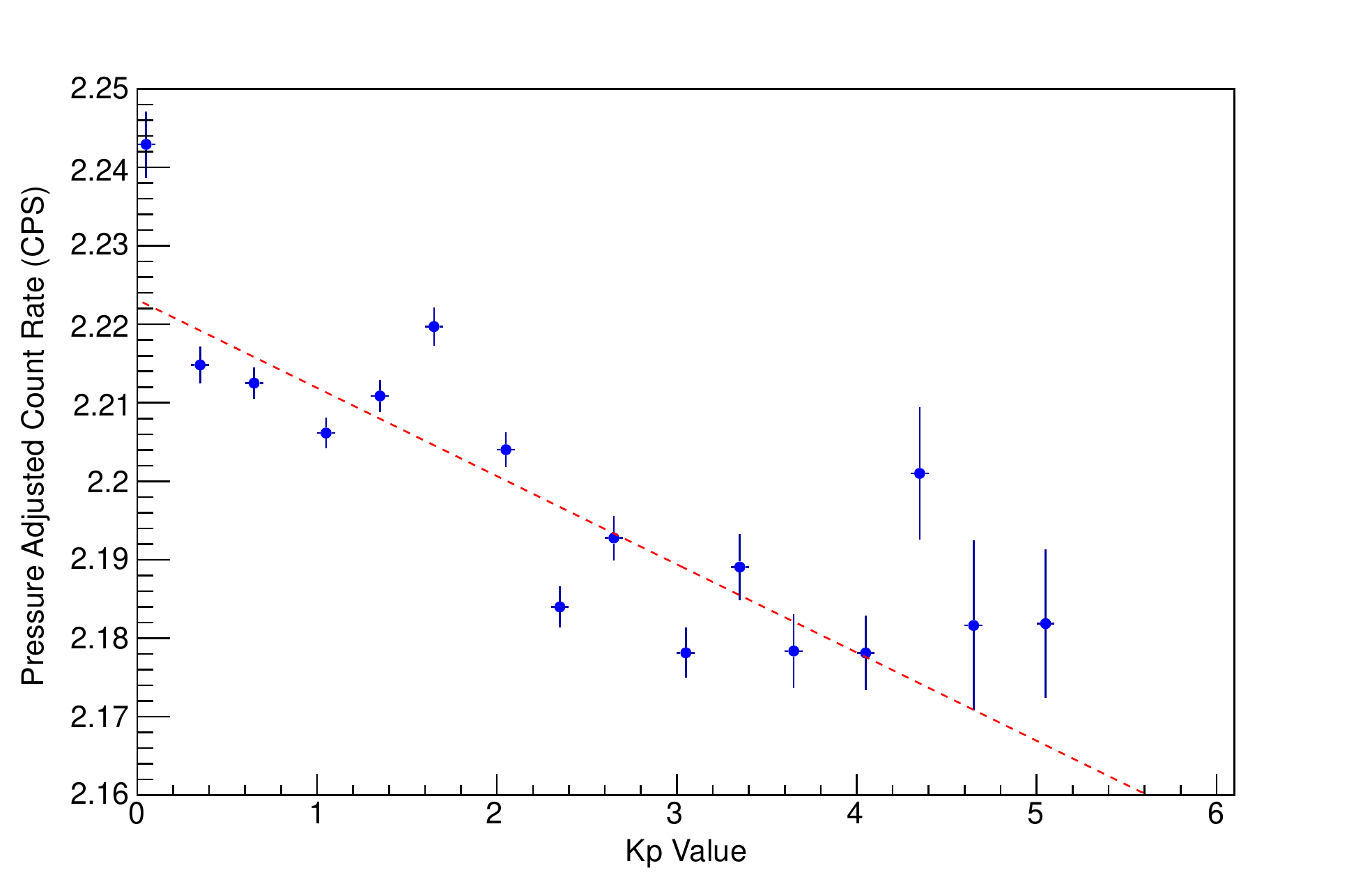}
		\caption{The pressure adjusted count rate given the $K_p$ index for all RadMAP data with linear fit $y=(-0.0112\pm0.0007) x+(2.223\pm0.001)$.}
		\label{fig:Kp_CR_Adj}
	\end{spacing}
\end{figure}
As expected, a suppression in the cosmic ray neutron count rate is observed with increasing magnetic field activity. Likewise, suppressed count rates are typically measured in ground based neutron monitors with increasing $K_p$ index. For the RadMAP data, a suppression of 2.3\% in the rate is observed at a $K_p$ index of 5 ($K_p$ index of 5 is classified as a minor geomagnetic storm). This small suppression is not significant enough to make a rate adjustment on data given that statistical and systematic uncertainties are generally much greater than 2.3\% (depending on the averaging time). 
However, it would still be important to be aware of significant GLE events or Forbush decreases that may affect the measured background rate for any neutron counting experiment in the field.

\section{Surrounding structures influence}\label{sec:SVF}

\begin{figure}[h!]
	\begin{spacing}{1.0}
		\centering
		\includegraphics[scale=0.38]{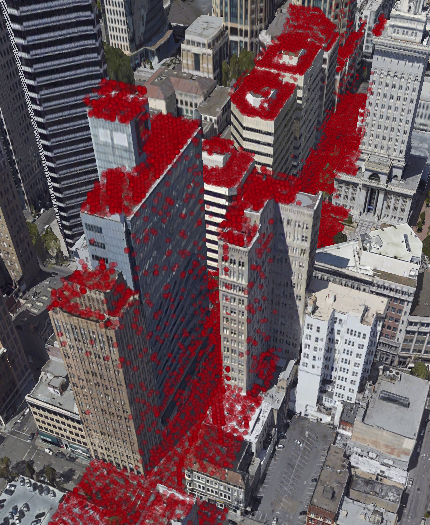}
		\caption{Overlay of USGS lidar data on a portion of 14th Street in Oakland in Google Earth's 3D structures view. Imagery $^{\footnotesize \copyright}$Google.}
		\label{fig:Lidar14th}
	\end{spacing}
\end{figure}

Using available National Oceanic and Atmospheric Administration (NOAA) coastal area lidar data~\cite{NOAA}, an algorithm was implemented to approximate sky-view factors (the total fraction of visible sky) for points along RadMAP’s route.  
Figure~\ref{fig:Lidar14th} shows one out of every five of the lidar data points that cover a portion of 14th Street in Oakland. Each USGS lidar point has latitude, longitude and elevation information associated with it.

The first step in the analysis was to write an algorithm to determine a two-dimensional open sky angle for each position of the detection system, similar to the angles determined manually by Iyengar et al.~\cite{Iyengar_Urban_Det}. In calculating this two-dimensional angle, we consider only the plane transverse to the truck's direction of motion.
In the algorithm, the lidar points must be ``scanned'' on either side of the truck to determine the height above the ground and the horizontal distance from the truck center. The horizontal distances are computed by determining the straight line distance between the truck's GPS coordinate and each of the lidar point coordinates.
The height of each lidar point is simply determined by subtracting the truck's GPS elevation value from the lidar elevation since the lidar elevations are also relative to sea level. In this study, only points at or above the truck's elevation are considered in the calculations so a maximum angle of 180$^{\circ}$ may be obtained.

\begin{figure}[h!]
	\begin{spacing}{1.0}
		\centering
		\includegraphics[scale=0.32]{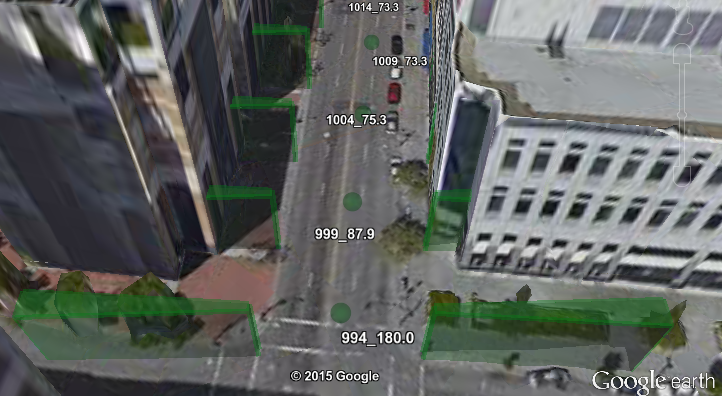}
		\caption{Select samples of the two-dimensional open sky angle search windows plotted on Google Earth. The numbers correspond to the bin number followed by the calculated open sky angle for the point. For example, the point labeled 999\underline{{ }{ }}87.9 is bin number 999 with an open angle of 87.9$^\circ$. Both the points and polygons are drawn elevated above the ground level for ease of viewing the overlay.}
		\label{fig:2D_OA_Google}
	\end{spacing}
\end{figure}

Three meter bins are defined by GPS coordinates along the path of the truck starting when the truck enters a specific geographic region defined for analysis. 
In calculating the two-dimensional open sky angle and the sky-view factor, search windows are defined based on the calculated bearing of the truck within the bin of interest.  
For clarity, only one-fifth of the location bins (every 15 meters rather than the actual 3 meter intervals) in which open sky angles are calculated are represented in Fig.~\ref{fig:2D_OA_Google}. Note that the green search windows as drawn have an artificial vertical extent which is only for ease of viewing and reducing clutter in the KML image. The actual search windows consider all lidar points at all elevations within the polygons as they are constrained only by angular and horizontal distance parameters.

Referring back to Fig.~\ref{fig:2D_OA_Google}, the road intersection at point 994 gives the proper result of a 180$^\circ$ two-dimensional open sky angle using the adjacent search windows given the truck's bearing. The glaring issue is this intersection has tall buildings located on each corner which in reality significantly reduce the portion of visible sky for that coordinate. Thus, the two-dimensional open sky angle calculation does not represent the environment under these conditions encountered by RadMAP. 
The sky-view factor is used as the basis for an alternative method to shading an image as a relief visualization technique in digital elevation model (DEM) generation. Zak\v sek et al. computes the sky-view factor for points on a surface to obtain greater detail of smaller features in relief images~\cite{Zaksek_sky_view}. Zak\v sek's calculation of the sky-view factor uses vertical elevation angles above the horizon at a fixed radius, $R$, from the point of interest.
The sky-view factor (SVF) is determined by Eq.~\ref{eq:SVF}.
\begin{equation}\label{eq:SVF}
\text{SVF} = 1-\frac{\sum_{i=1}^n\text{sin}\gamma_i}{n}
\end{equation}
Where $\gamma_i$ is the elevation angle from the horizon to the highest obstructed point in slice $i$ and $n$ is the number of slices used to estimate the sky-view factor at the point of interest along the route.

\begin{figure}[h!]
	\begin{spacing}{1.0}
		\centering
		\includegraphics[scale=0.35]{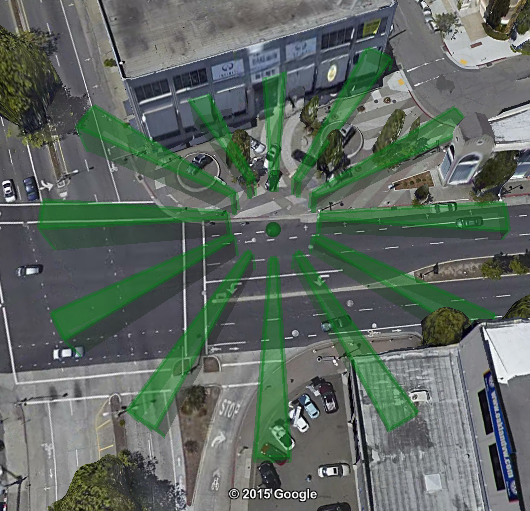}
		\caption{The 12 lidar search windows used to compute the sky-view factor for one point along RadMAP's path. Imagery $^{\footnotesize \copyright}$Google.}
		\label{fig:SVF_Slices}
	\end{spacing}
\end{figure}

In our study, the sky-view factor for every 3 meters of RadMAP's travel was calculated using Zak\v sek's method. Instead of only considering the two search windows as used for the two-dimensional open sky angle method, 12 open sky angles were calculated for windows in 30$^\circ$ intervals around the truck's position. A 35 meter radius around the truck was used to search for lidar points. To obtain the elevation angles, each of the 12 open sky angles is subtracted from 90$^\circ$ and the sky-view factor is calculated using Eq.~\ref{eq:SVF} with $n=12$. The search windows are illustrated in Fig.~\ref{fig:SVF_Slices}.
Additional angular samples ($n>12$) could be taken, however, the trade-off is in the additional computing time it takes to determine the elevation angles when searching through the large number of lidar data points.

\begin{figure}[h!]
	\begin{spacing}{1.0}
		\centering
		\includegraphics[scale=0.48]{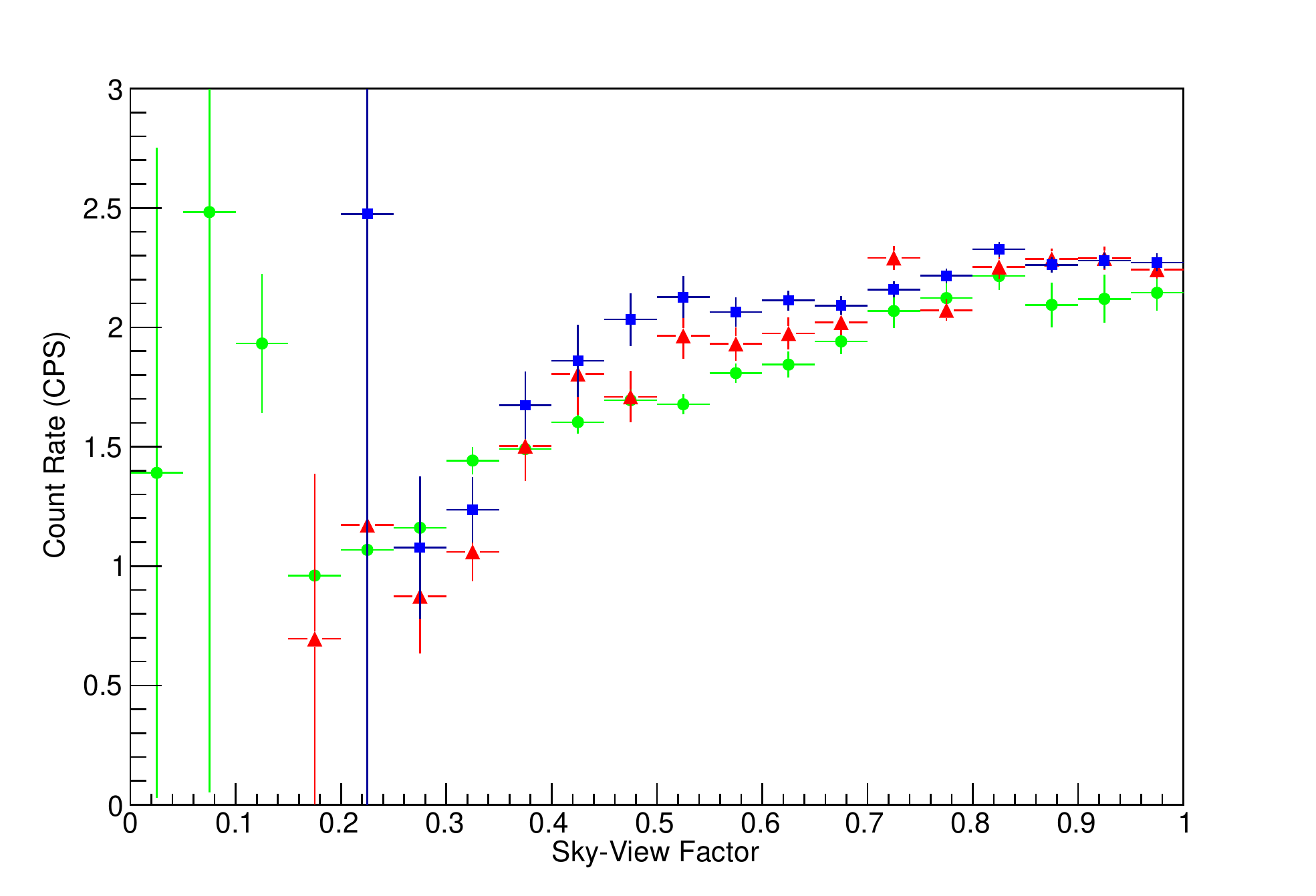}
		\caption{Pressure adjusted count rates given sky-view factors plotted separately for Berkeley (blue squares), Downtown Oakland (red triangles), and Downtown San Fransisco (green circles). Large error bars on data points represented by low statistics are cut off to increase the ease of distinguishing data points.}
		\label{fig:SVF_CR_Separates}
	\end{spacing}
\end{figure}

The result for pressure adjusted count rates given sky-view factors for each of the three areas studied is presented in Fig.~\ref{fig:SVF_CR_Separates}.
The three downtown areas are fairly self-consistent with the amount of suppression at lower sky-view factors. Downtown Berkeley seems to exhibit slightly higher count rates across the board and downtown San Fransisco seems to yield slightly lower count rates especially at higher sky-view factors. This difference could be due to the density of tall buildings in San Fransisco versus Berkeley. 
This is likely due to an overestimation of the sky-view factor in San Francicso given the greater density of tall buildings outside the 35 meter radius cylinder used for the calculation. Studies with a greater radius may be conducted to confirm this, however, computing time will increase significantly with an increase in radius.

\begin{figure}[h!]
	\begin{spacing}{1.0}
		\centering
		\includegraphics[scale=0.49]{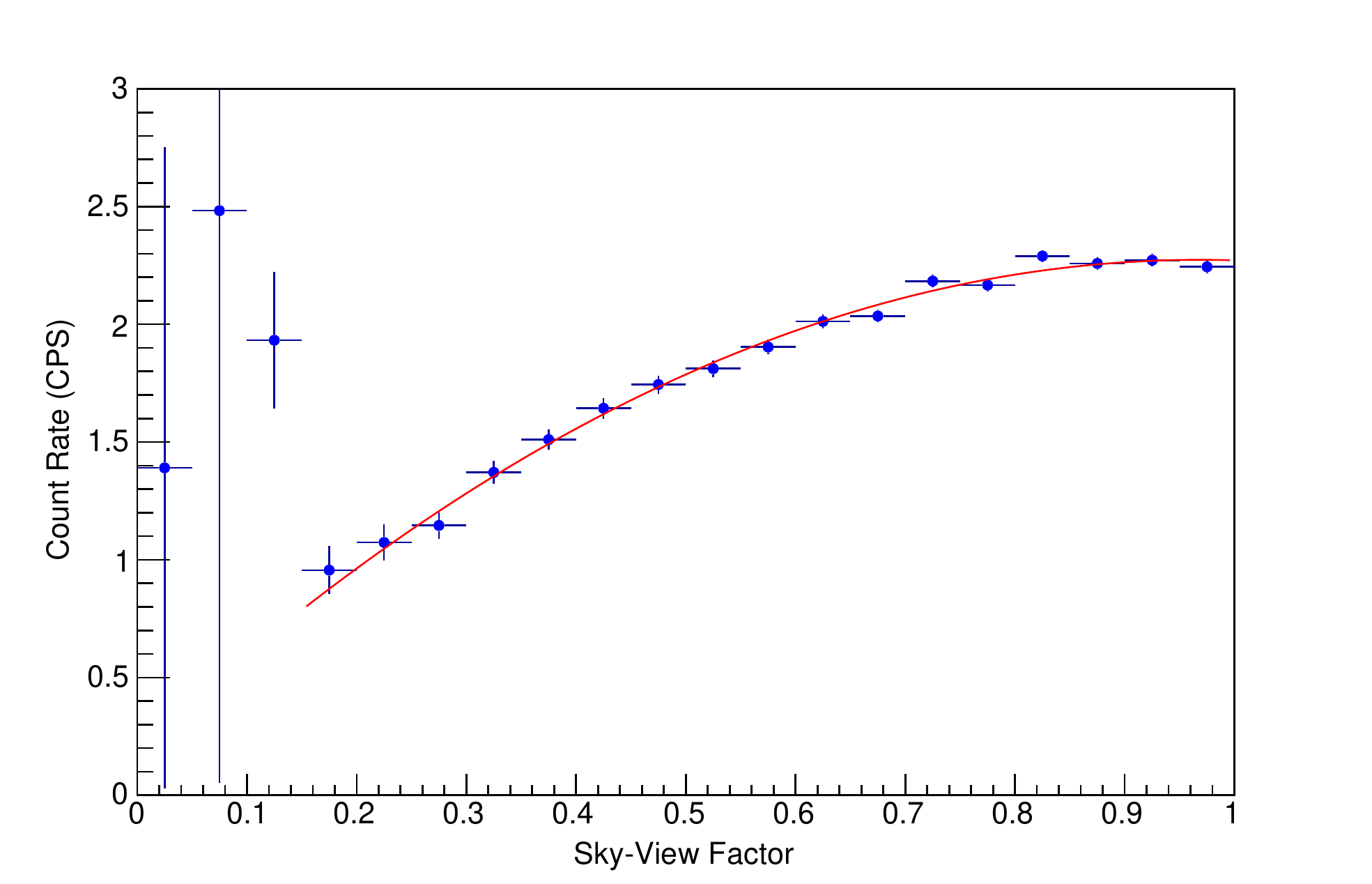}
		\caption{Pressure adjusted count rates for sky-view factors from all runs combined that travel through select portions of Berkeley, Downtown Oakland, or Downtown San Fransisco fitted with the polynomial function $y=(-2.2\pm0.2)x^2+(4.3\pm0.2)x+(0.19\pm0.08)$. Large error bars on low sky-view factors with poor statistics are cut off for ease of viewing.}
		\label{fig:SVF_CR_All}
	\end{spacing}
\end{figure}

Figure~\ref{fig:SVF_CR_All} shows the combined result for sky-view factor fitted with a quadratic function. 
The bin representing sky-view factors from 0.1 to 0.15 shows an anomalously high count rate compared to the trend. Further investigation revealed that most of the events in this bin came from a particular location in downtown San Francisco. At the time of the USGS lidar dataset acquisition, this location contained densely packed tall buildings and overhead structures, so the low SVF was correctly calculated from the lidar data. However, these buildings and structures were demolished prior to the RadMAP data runs, so that the true SVF at the time of neutron data acquisition was much higher.  Due to this discovery and the uncertainty at lower sky-view factors, we fit to values greater than 0.15 in Fig.~\ref{fig:SVF_CR_All}.

The anomaly discovered at the downtown San Francisco location with unrepresentative lidar data reinforces the predictive value of the SVF, but clearly shows the disadvantage of a temporal gap between the lidar data and neutron data acquisition, which is inherent in the use of independent USGS lidar data.  A preferred approach would be to use onboard lidar to determine the SVF at the same time as neutron data acquisition. Although RadMAP does collect lidar data, its field of view does not include overhead angles, so further hardware additions or modifications are needed in order to test this approach using RadMAP.

Between the resulting pressure adjusted count rate of 2.25 CPS at a sky-view factor of 1 and a rate of 0.95 CPS at a sky-view factor of 0.2 a suppression of 58\% in the rate was observed. The result obtained tells us the shielding effect of buildings in urban environments is dominant over any additional production in building materials from spallation processes. The suppression at low sky-view factors is the greatest residual effect (after pressure adjustment) studied and represents a significant influence on the neutron background count rate. For comparison, the pressure effect yielded a suppression of about 32\% from 970 mbar to 1030 mbar. In urban area search applications, an adjustment could be made given the sky-view factor in addition to the pressure to increase detectability of sources over background.

\section{Conclusions}\label{sec:Conclusions}

The fast neutron background characterization studies in this paper both complement and enhance ongoing and previously conducted research in this field. Data obtained by the organic liquid scintillator cell system on RadMAP exhibited good agreement with observations originally made by Pfotzer on background event rate at various altitudes. Pressure adjustments applied to detected events effectively reduce systematic error contributions to the overall background count rate distribution. The reduction in background uncertainty may increase the detectability of neutron emitting sources, a critical goal for SNM detection. Results also complement and extend current research on the suppression of the fast neutron count rate in urban areas. This study employed a novel method and added a layer of complexity to previously published research by using urban lidar data and the calculation of the sky-view factor to characterize the magnitude of background suppression. 
This paper provides a comprehensive characterization of the fast neutron background and employs methods that may be applied to various detection scenarios and systems, both mobile and stationary. The end-state of such an in-depth characterization of the neutron background is the improved detectability of neutron-emitting material and SNM.

\section*{Acknowledgement}

Special thanks to Jim Brennan and Dan Throckmorton from SNL for mounting and installing the liquid scintillator cells. Bryan Clifford, Aaron Nowack, and Thorin Duffin from SNL were responsible for setting up the data acquisition system, initial stages of analysis, and data quality monitoring.
This work was also made possible by the hard work and dedication of the entire RadMAP crew and research team at LBNL including Mark Bandstra, Timothy Aucott, Victor Negut, Joseph Curtis, and Ross Meyer.

Sandia National Laboratories is a multi-mission laboratory managed and operated by Sandia Corporation, a wholly owned subsidiary of Lockheed Martin Corporation, for the U.S. Department of Energy’s National Nuclear Security Administration under contract DE-AC04-94AL85000.

\bibliographystyle{ieeetr}
\bibliography{N_Background_References/n_bibtex}

\end{document}